\documentclass[runningheads]{llncs}
\usepackage[T1]{fontenc}
\usepackage{graphicx}
\usepackage{booktabs}
\usepackage{booktabs}
\usepackage{amsmath}
\usepackage{algorithm}
\usepackage{algorithmic}
\usepackage[switch]{lineno}
\usepackage{multirow}
\usepackage{hyperref}
\usepackage{longtable}

\usepackage[misc]{ifsym}
\newcommand{\corr}{(\Letter)}
\usepackage{mwe}
\begin{document}
%\title{Bias vs Bias: Dawn of Justice - A Category-aware Re-ranking Approach for Fair Recommendations}

% \title{Bias vs Bias: Dawn of Justice}
%\title{Bias vs Bias: A Fair Fight in Recommendation Systems}
\title{Bias vs Bias - Dawn of Justice:\\ A Fair Fight in Recommendation Systems}
\titlerunning{A Fair Fight in Recommendation Systems}

% \titlerunning{Underwater Basket Weaving Under Extreme Pressure}
% If the full title of your paper is short enough to also fit in the running head, you can omit the abbreviated paper title here. You can check as follows: if you comment out the \titlerunning line, something will appear in the header of all odd-numbered pages of your PDF from page 3 onward. This something is either the full title (in which case all is well), or the error message "Title Suppressed Due to Excessive Length". If this error message appears, you're going to want to provide an abbreviated title within the \titlerunning command, because if you won't do it, Springer will do it for you.

%N.B.: Author information (both in the \author{} and \authorrunning{} command) should only be present in the Camera-Ready Version of your paper. The version that you initially submit for review, ought to be double-blind. So, when initially submitting your paper, use:
\author{Author information scrubbed for double-blind reviewing}

\author{Tahsin Alamgir Kheya\corr \orcidID{0009-0001-2481-2877}  \and
Mohamed Reda Bouadjenek \orcidID{0000-0003-1807-430X} \and
Sunil Aryal\orcidID{0000-0002-6639-6824}}
\tocauthor{Tahsin Alamgir Kheya, Mohamed Reda Bouadjenek, Sunil Aryal }
\toctitle{Bias vs Bias - Dawn of Justice:\\ A Fair Fight in Recommendation Systems}
% You may leave out the orcidID information, if you want to.
% Use \corr to indicate the corresponding author. Note the spacing around the \corr command. Only one author can be the corresponding author.

%N.B.: comment out the \authorrunning{} command for the double-blind version of your paper submitted for review. Later, if your paper is accepted, use the command for the Camera-Ready Version.
% \authorrunning{T.A. Kheya et al.}
% If there are two authors, write 'A.L. Benjamin and C.C. Broadus Jr.'
% If there are more than two authors, '[...] et al.' is used.

\institute{Deakin University, Geelong, VIC, Australia \email{\{t.kheya,reda.bouadjenek,sunil.aryal\}@deakin.edu.au}}

\maketitle              % typeset the header of the contribution
\textit{This paper has been accepted to the ECMLPKDD2025 conference.}
\begin{abstract}
Recommendation systems play a crucial role in our daily lives by impacting user experience across various domains, including e-commerce, job advertisements, entertainment, etc. Given the vital role of such systems in our lives, practitioners must ensure they do not produce unfair and imbalanced recommendations. Previous work addressing bias in recommendations overlooked bias in certain item categories, potentially leaving some biases unaddressed. 
Additionally, most previous work on fair re-ranking focused on binary-sensitive attributes. 
In this paper, we address these issues by proposing a fairness-aware re-ranking approach that helps mitigate bias in different categories of items. This re-ranking approach leverages existing biases to correct disparities in recommendations across various demographic groups. We show how our approach can mitigate bias on multiple sensitive attributes, including gender, age, and occupation. We experimented on three real-world datasets to evaluate the effectiveness of our re-ranking scheme in mitigating bias in recommendations. Our results show how this approach helps mitigate social bias with little to no degradation in performance.

\end{abstract}

\section{Introduction}
Recently, Recommendation Systems (RSs) have become an integral part of our lives by providing personalized suggestions to us. They play an important role in shaping our digital experience, contributing to our decisions for online purchases, movie recommendations, music playlists, news feeds, and more. RS spares us the trouble of sifting through vast amounts of data by curating customized and diverse content. Given their profound impact on our daily lives, it is essential to ensure they provide fair recommendations and do not perpetuate harmful biases. For instance, \cite{10.1145/3132847.3132938} highlights how top-ranked results for job roles favor one gender over the other and systematically disadvantage minority groups. While significant progress has been made in addressing fairness in recommendations \cite{10.1145/3531146.3533238,10.1145/3460231.3473897,wang2023survey}, it is still an ongoing topic of research with new studies emerging continuously. 

\begin{figure}[!htb]
    \centering
    \includegraphics[width=1\textwidth]{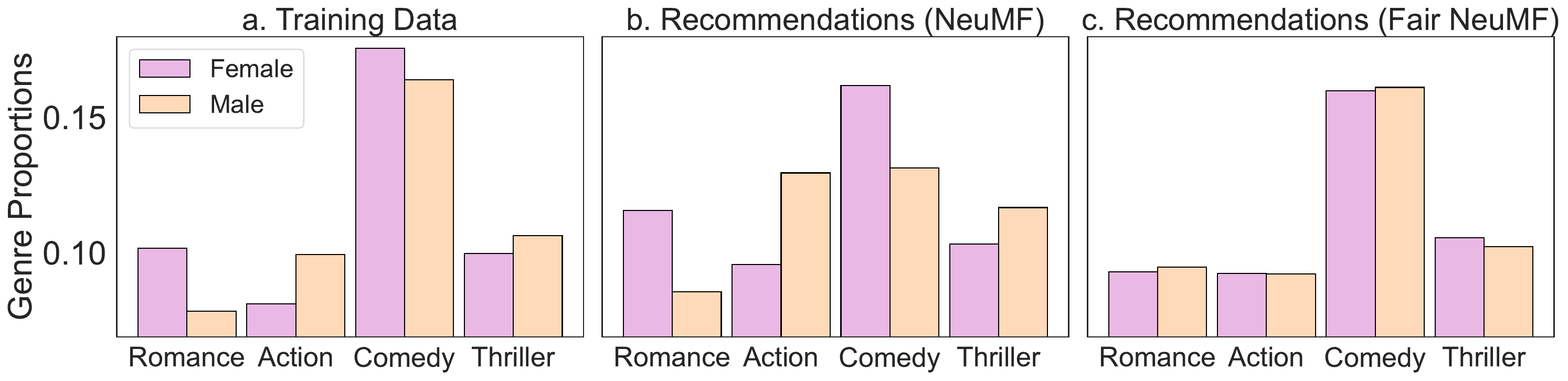}
    \caption{
    Comparison of proportions of different movie genres for users of two genders in the training data, plain recommendations, and fairness-aware recommendations. There are disparities in the number of movies recommended to each gender for the four genres. This graph is based on the NeuMF \cite{10.1145/3038912.3052569} model for the ML100K dataset \cite{10.1145/2827872}. }
   
    \label{fig:compare_genre}
    % \vspace{-0.5cm}
\end{figure}

AI models are vulnerable to picking up biases that exist in the dataset used to train them \cite{10.5555/3157382.3157584,lum2016,pmlr-v81-ensign18a,10.1145/3287560.3287572,kheya2024pursuitfairnessartificialintelligence}. For instance, \cite{alamgir2025unmasking} discusses how there is significant bias in movie recommendations for male and female users, across genres like romance and action. 
To mitigate such biases, researchers have used various fairness constraints to design re-ranking algorithms \cite{10.1145/3132847.3132938,Fu2020,10.1145/3292500.3330691,pmlr-v139-gorantla21a}. While there has been abundant work in the field of fair re-ranking in recommendations, existing approaches are deficient in two key ways: (i) they primarily focus on single or binary sensitive attributes and, (ii) they do not include the item categories when designing their approaches, which plays a vital role in users' end experience. 
\\ \hfill

To illustrate the importance of mitigating bias on a granular scale by considering categories, we refer to Figure \ref{fig:compare_genre}.  
In this figure, we present how the proportions of movies from different genres vary for male and female users using the ML100K dataset across three stages: the training set, the top 20 plain recommendations provided by the NeuMF model, and the top 20 recommendations after fair re-ranking. From this figure, we can derive and explain several key insights: (i) The dataset used to train the model is not fair or neutral as shown in plot a. Certain genders tend to exhibit inherent biases towards specific categories, which societal stereotypes can influence. (ii)  During the training process, the model can learn these biases and amplify them, intensifying their impact on the final output. We choose to demonstrate this using one popular recommendation model called NeuMF (plot b). The amplification of such biases by recommendation models is very common, as noted by \cite{lin2019crankvolumepreferencebias,mansoury2019biasdisparitycollaborativerecommendation}. While this helps visualize the discrepancy in movie recommendations for different genders, similar biases can exist in other domains, such as news recommendations, which can have far more profound consequences if not addressed. A similar concern is highlighted by \cite{yesilada2022systematic}, where YouTube recommender systems are found to facilitate pathways to extremist or radicalizing content. These kinds of content, when exposed to users, especially young generations, may have negative impacts on their well-being. (iii) The biases present in plots a and b, have similar trends of being oppositely skewed. As such, male users are more biased towards action and thrillers, while less towards romance and comedies. And female users are more biased towards romance and comedies while less towards action and thrillers. We take advantage of these opposing biases and use them as a corrective mechanism against social bias. After applying our fair category-aware re-ranking approach the discrepancy in the categories of movies recommended to the different groups of users decreases significantly as seen in plot c.

Ensuring the categories of items are considered when designing a re-ranking scheme that also caters to multi-valued sensitive attributes is thus vital. This kind of refined approach will ensure a balanced distribution of recommended categories of items for different groups of users. 
In this paper, we introduce a fairness-aware re-ranking scheme that allows us to produce fair recommendations by considering users' social attributes, accommodating both binary and multi-valued attributes. This strategy builds on the concept of counterfactual fairness by leveraging the bias in the training set to tackle/counteract social bias in a category-aware setting. Essentially, we use the category preferences of users with different sensitive attributes to adjust the recommendations, leveraging the opposing biases to promote fairness. We evaluate the effectiveness of this scheme on different recommendation algorithms (including traditional and deep approaches), experimenting on three real-world datasets. 

% The contribution of this paper can be summarized as follows:
% \begin{itemize}
%     \item We introduce a new re-ranking system to produce recommendation lists for users which mitigate consumer-side bias.
% \end{itemize}
% \subsection{Main Contributions}
% Please note that the first paragraph of a section or subsection is
% not indented. The first paragraph that follows a table, figure,
% equation etc. does not need an indent, either.

% Subsequent paragraphs, however, are indented.

\section{Related Work}
\subsection{Consumer-Side Fairness in Recommendation Systems}
Fairness in recommendation systems is a multi-sided concept, which is categorized into (i) Provider-Fairness: fairness for providers or sellers in terms of exposure; (ii) Consumer-Fairness: which focuses on the fairness of items being recommended to users from different protected classes and (iii) CP-Fairness which considers both \cite{burke_multisided_2017}. Our work focuses on C-Fairness, with our goal of similar recommendations regardless of the user's sensitive attributes. It has been observed in prior research how recommendation systems are prone to bias influenced by demographic factors like gender \cite{edizel_fairecsys_2020,neophytou2021revisiting,MELCHIORRE2021102666,pmlr-v81-ekstrand18b,article}, age \cite{neophytou2021revisiting,pmlr-v81-ekstrand18b}, occupation \cite{10.1145/3404835.3462966,10.1145/3442381.3450015}, race \cite{10.1145/2447976.2447990} and more. To promote fairness for consumers, researchers have proposed a variety of mitigation strategies that span across the pre-processing, in-processing, and post-processing stages of the ML pipeline. For instance, \cite{10.1145/3289600.3291002} shows how small additions of augmented data can substantially improve both individual and group fairness in recommender systems. The authors in \cite{10.1145/3534678.3539269} propose a
multi-task adversarial learning scheme that satisfies three different fairness criteria, including group, individual, and counterfactual. Optimizing a fairness-aware regularization term along with the main recommendation loss is also a popular approach to mitigating bias in recommendations \cite{Wu_Wu_Wang_Huang_Xie_2021,10.1145/3292500.3330745,Boratto21,alamgir2025unmasking,NIPS2017_e6384711}. 
\subsection{Fair Re-ranking}
Re-ranking is a popular post-processing strategy to mitigate bias in recommendations. This method focuses on rearranging the items recommended to users for the top-$k$ list by considering both recommendation quality and a fairness constraint. For instance, \cite{Fu2020} introduces a re-ranking approach that incorporates a fairness constraint to mitigate unfairness in explainable recommenders that use knowledge graphs. The authors in \cite{10.1145/3442381.3449866} introduce a fairness-constrained re-ranking method to ensure the utility disparity between different groups of users are below a certain threshold \(\epsilon\), while the optimization maximizes preference scores of items selected. Singh and Joachims \cite{10.1145/3219819.3220088} integrate
common fairness concepts like demographic parity, disparate impact, and disparate treatment into their optimal ranking algorithm. The main idea behind the most relevant work is optimizing fairness and utility jointly by using a hyper-parameter to control the trade-off \cite{wang2023survey}. Although the current research community has not explored fair re-ranking for multi-valued attributes as much, there are a few works we wanted to highlight \cite{10.1145/3397271.3401292,NEURIPS2022_cdd06402,DBLP:journals/corr/abs-2006-08688,hua-etal-2024-up5}. 
Unlike these, we are enforcing C-Fairness for multi-valued attributes in a category aware-setting.
Most existing fair ranking schemes for recommendations focus on binary sensitive attributes and apply fairness definitions using the intuition of Equalized odds and Demographic parity to design their fairness constraint \cite{10.1145/3442381.3449866,10.1145/3219819.3220088,10.1145/3477495.3531959,10.1145/3292500.3330745}. For demographic parity each sensitive group (like male and female) should receive the same proportions of positive predictions \cite{hardt2016equality}. On the other hand, the concept of Equalized odds holds if the system has similar true positive rates and false positive rates across two different demographic groups \cite{hardt2016equality}. In the current literature, these works would try to minimize the disparity between two groups of users or items based on popularity or user-sensitive attributes with binary values (like binary gender: [male, female], or age: [old, young]). While this approach is valuable in some way, it tends to oversimplify the complexity of user identities that are multi-dimensional. 
% However there are some works which still focus on multiple groups when reranking for provider-side\cite{10.1145/3397271.3401292,NEURIPS2022_cdd06402} fairness and 

Additionally, most re-ranking schemes don't consider the proportion of categories in the items recommended. Such schemes can fail to mitigate bias and disparities that exist across different types of items (like genres for movies). There are however some works that do consider different classes when designing ranking schemes \cite{10.1145/3292500.3330691,10.1145/3240323.3240372,10.1145/3626772.3657794,pmlr-v139-gorantla21a}. For instance, \cite{10.1145/3292500.3330691} introduces algorithms to re-rank job candidates to achieve a desired distribution in the top results in regards to users' sensitive attributes like gender and age. The works by  \cite{10.1145/3626772.3657794,pmlr-v139-gorantla21a,10.1145/3292500.3330691} aim for provider-side fairness. Our work is very close to that of \cite{10.1145/3240323.3240372}, which re-ranks movies to ensure the recommendations align well with the historical interaction of the users by using genre distributions of previously played movies. Although this work is generating fair recommendations by ensuring users get recommendations following the proportions of genres that they previously watched, our work focuses on fairness in terms of users' sensitive attributes.

% Our ranking scheme addreses these gaps by coming up with an idea that can handle both binary and multinary sensitive attributes. Plus too ensure fairness is achieved on a granular level, we consider the category distibutions of items recommended. 

% \section{Motivating Category Bias Concern in Recommendation Systems}
% To explain the importance of mitigating bias in recommendations in a granular level we provide a motivating example. Our example is a typical offline setting recommender system. This recommendation system is trained on historical hiring data, which is often biased due to societal norm. Following these stereotypes, the recommender system can discriminate against users based on their sensitive attributes. For instance male users might receive more leadership roles, whereas women receive more caregiving roles. When designing mitigation strategies for such biases, the usual approach is to ensure user groups of different sensitive attributes are getting similar recommendations according to their utility scores or performance metrics. While simple and straight-forward these methods can fail to mitigate bias that exists among the different categories of items recommended. 

\begin{table}[!htb]
\centering
\caption{Notation Table}
\label{tab:notation}
\resizebox{1\columnwidth}{!}{%
\begin{tabular}{|l|p{12cm}|}
\hline
\textbf{Notation} & \textbf{Description} \\ \hline
           % $u_i$     &   A single user, where $i$ indexes the users.                   \\ \hline
             
             $\mathcal{U}$ and $\mathcal{V}$   & The set of users and items, respectively.   \\ \hline
              $v_j$  &   A single item, where $j$ indexes the items.                   \\ \hline
               $c$ &        An item category, such as Action, Sci-Fi, Romance, etc.              \\ \hline
               
               $\mathcal{C}$ & The list of unique categories associated with all items. \\ \hline
               $C$ & A category matrix where $C_{v,c} = 1$ if item $v$ belongs to category $c$, and $0$ otherwise.                \\ \hline
              $C_{v}$ & The list of categories associated with item $v$.                     \\ \hline
            % $TopK_{u}$ & The set of top $K$ recommended items for user $u$.                         \\ \hline
            % $score_{u,v}$ & Is the predicted score for item $v$ by user $u$. \\ \hline
              $\mathcal{V}_{u}$ & The set of items the user \(u\) has  interacted with in the past. %In our case, this would be the training set of user \(u\).                       
                \\ \hline
               $t_{v,u}$& The timestamp of the interaction with item \(v\) by user \(u\).                      \\ \hline
               $s_u$ & Represents the value of a sensitive attribute (male, female, engineer, etc.) for user $u$. \\ \hline
               $S$ & Represents a sensitive attribute like age, gender, occupation, etc. \\ \hline
                $\mathrm{score}_{u,v}$ & The predicted score of item \(v\) by user \(u\). \\ \hline
\end{tabular}}
\end{table}

\section{Proposed Re-ranking Scheme}

\subsection{Notation}
We present all metrics for fairness assessment using the mathematical notation presented in Table \ref{tab:notation}.

We start with the two main distributions, both of which consider categories of the items. 
\begin{definition}[Counterfactual Category Proportion (CCP)]
Let \( o(c|s_u) \) return the average proportion of category $c$ for all users who have a sensitive attribute that is not \(s_u\), where \(\mathcal{U}_{\neg s_u} = \{ w \in \mathcal{U} \mid s_u \neq s_w \}\).
\end{definition}

\begin{equation}
    o(c|s_u) =  \frac{1}{|\mathcal{U}_{\neg s_u}|} \sum_{{u \in 
 \mathcal{U}_{\neg s_u}}} m(c|u)
\label{eq:p}
\end{equation}
where 
\[m(c|u) =  \frac{\sum_{v \in \mathcal{V}_{u}} \frac{C_{v,c}}{|C_{v}|} \cdot t_{v, u} }{\sum_{v \in \mathcal{V}_{u}}t_{v,u}}\]

We use the timestamps of the interactions to apply more weight to interactions that took place recently. We follow \cite{10.1145/3477495.3531959}, where they also employed the training set for their re-ranking algorithm.
\begin{definition}[Recommended Category Proportion (RCP)]
Let \(r(c|u)\) be the category proportion for user \(u\) relative to the items they are being recommended (represented by I) for category \(c\).
\end{definition}
      \begin{equation}
    r(c|u,I) =  \frac{\sum_{j=1}^{|I|} \frac{C_{v_j,c}}{|C_{v_j}|} \cdot \frac{1}{j^{\gamma}} }{\sum_{j=1}^{|I|}{\frac{1}{j^{\gamma}}}} 
\label{eq:r}
\end{equation}
Where we use \(\gamma \in [0,1]\) to help us weigh the item category contribution according to the rank ($j$) of the item in the recommended list $I$%the top-$k$ list.  

\subsection{Counterfactual Fairness}
To design our category-aware fair re-ranking scheme, we use the concept of counterfactual fairness \cite{NIPS2017_a486cd07} that is formally defined as:
\begin{equation}
    P\big(\hat{Y}_{A \gets a}(U) = y \mid X = x, A = a\big) 
\nonumber \\
= P\big(\hat{Y}_{A \gets a'}(U) = y \mid X = x, A = a\big)
\end{equation}
    Here, counterfactual fairness is achieved if the predicted outcome \(\hat{Y}\) for an individual $u$ (with latent variable \(U\) and non-sensitive features $X$) is the same when intervening to externally set the user's sensitive attribute from \(a\) to \(a'\). This concept avoids discrimination by making sure that sensitive attributes do not influence the outcomes unfairly.
    Using this intuition, we extend it by not only considering individual-level outcomes but also including a group-level distribution of historical interactions for a fairness reference point.
\subsection{Proposed Fair Re-ranking Idea}
When designing our re-ranking approach, we want to adjust the recommendations based on how users of different sensitive groups interact with items of different categories. For this, we leverage the popularity of different categories among users with different sensitive attributes from their historical interactions. By doing so, we effectively simulate a counterfactual scenario, where we use category preferences of users who do not share the same sensitive attributes. To achieve this we want to ensure that the deviation between:
\begin{itemize}
    \item the category distribution recommended to a user, where the proportion of a single category is defined by \(r(c|u,I)\) (as shown in Equation \ref{eq:r}) and
    \item the average category distribution of users who don't share the same sensitive attribute as this user, where the proportion of a single category is defined by \(o(c|s_u)\) (as shown in Equation \ref{eq:p}) 
is minimized. 
\end{itemize}
Essentially, \(o(c|s_u)\) acts as a counterfactual baseline for us that helps counteract the tendency of recommenders to reinforce
existing biases from the data they are trained on. This intuition will help align users' recommendations and act as a defying
mechanism for historical bias.

To quantify the disparity between the two distributions, we will use KL divergence. Using Kullback-Leibler (KL) divergence, in this case, has numerous advantages, such as sensitivity to subtle differences in the two category distributions, alignment with counterfactual definition, where we capture the difference in how recommended items differ when sensitive attributes are changed and ease of interpretation. The equation below helps quantify the disparity between the two distributions:
\begin{equation}
    D_{KL}(o||r(I)|u) = \sum_{c \in \mathcal{C}} o(c|s_u) log \frac{o(c|s_u)}{\tilde{r}(c|u,I)}
\end{equation}
where 
\[\tilde{r}(c|u,I) = (1-\alpha) \cdot r(c|u,I) + \alpha \cdot o(c|s_u)\]
To avoid getting any value of \(r(c|u,I) = 0\), we use \(\tilde{r}\) where \(\alpha\) is a really small number between 0 and 1. Note that here, $o$ and $r(I)$ represent the distribution of CCP and RCP across all categories for user \(u\).
% the average category distribution of items interacted by users who don't identify as \(s_u\) and the category distribution of items recommended to \(u\) respectively.

We use an adaptation of Maximum Marginal Relevance (MMR) \cite{10.1145/290941.291025} to determine the optimal set of items \(I^*\), which can be formalized as:
\begin{equation}
I^* = \text{argmax}_{I \subseteq TopN, |I| = k}  (1 - \beta) \cdot rel(I,u) - \beta \cdot D_{\text{KL}}(o, r(I),u) 
\label{eq:o}
\end{equation}

where 
\[rel(I,u) = \sum_{v \in I} \mathrm{score}_{u,v}\]
We use a hyperparameter \(\beta \in [0,1]\) to calibrate the trade-off between relevance and fairness like some previous works, including \cite{10.1145/3213586.3226206,10.1145/3298689.3347016,10.1145/3240323.3240372}. This gives us a combinatorial optimization problem that is NP-hard. Following the works by \cite{10.1145/3240323.3240372,6854375}, which demonstrated that the greedy optimization of an equation similar to Equation \ref{eq:o} is equivalent to the greedy optimization of a surrogate submodular function, we adopt a similar approach condensing our equation to:
% \begin{align}
\begin{equation}
I^* = \text{argmax}_{I \subseteq TopN, |I| = k}  (1 - \beta) \cdot rel(I,u)
\nonumber \\
+ \beta \cdot \sum_c o(c|s_u)log \sum_{j=1}^{|I|} \frac{1}{j^{\gamma}} \Tilde{r}(c|v_j)
\label{eq:go}
\end{equation}
where \(\tilde{r}(c|v_j) = (1-\alpha) \cdot r(c|v_j) + \alpha \cdot o(c|s_u)\), and represents the proportion of category \(c\) in movie \(v_j\). The simplified submodular greedy optimization has an optimal guarantee of \(1-\frac{1}{e}\) \cite{nemhauser1978analysis}. The algorithm for this optimization is presented as Algorithm \ref{alg:algorithm}. Here we generate the top \(N\) items for each user (represented by \(TopN\)) and then re-rank to find the top \(k\) items (where \(N\geq k\)). Instead of using \(o(c|s_u)\) directly, we add a small constant variation across all \(c\) values (the impact of which can be considered negligible) to ensure non-zero entries. Additionally, we normalize the relevance term and fairness term through the min-max normalization scheme to ensure they are on the same scale.
\\
We want to mention that although we aim to provide fair recommendations to the users based on their sensitive attributes, we ensure this does not come at the expense of personalization. For our fair scheme, the goal is still prioritizing the preferences of users, but in a way that prevents the reinforcement of social stereotypes. 
\begin{algorithm}[t]
    \caption{The Counterfactually Fair Re-ranking Optimization}
    \label{alg:algorithm}
    \textbf{Input}: \(\mathcal{U}, TopN, \beta, k, scores, S\)\\
        \textbf{Output}: Matrix \( R \) of size \( |\mathcal{U}| \times k \) which contains fair top-\( k \) recommendation lists for each user.

    \begin{algorithmic}[1] %[1] enables line numbers
        \STATE \(scores \gets\) train baseline model and store scores of candidate items.\  
        % \caption{Matrix Initialization}
\STATE  $R \gets \text{empty matrix of size } |\mathcal{U}| \times k$
% $\STATE \(R \gets \emptyset\) \COMMENT{Matrix of size \(  \times k \), initialized to store recommended items.}

        \STATE \(H(u) \gets\) store historical interactions of all users.
        \STATE Compute \(o\) for all possible $s_u$ values, following Equation \ref{eq:o} for chosen S
        \FORALL{users \( u \in \mathcal{U} \)}
            \FOR{index = 0 \textbf{to} $k$-1}
            \FORALL{items \( i \in TopN_{u} \setminus R(u) \)}
                \STATE Compute fairness-aware scores for \(i\) using:
                \[
                (1 - \beta) \cdot rel(I, u) + \beta \cdot \sum_c o(c|s_u) \log \sum_{j=1}^{|R(u) \cup i|} \frac{1}{j^{\gamma}} \Tilde{r}(c|v_j)
                \]
            \ENDFOR
            \STATE Select the item \(i^*\) with the highest fairness-aware score.
            \STATE Add \(i^*\) to \(R(u)\).
        \ENDFOR
        \ENDFOR
        \STATE \textbf{return} \(R\)
    \end{algorithmic}
\end{algorithm}

\section{Experimental Methodology}
\subsection{Datasets}
We evaluate the effectiveness of our scheme on three publicly available datasets from different domains as shown in Table \ref{tab:plain}. 
The datasets are all pre-processed to remove items and users by k-core filtering, which is a common practice adopted in prior research \cite{10.1145/3404835.3463245,alamgir2025unmasking,dietz_understanding_2025}. 
In our case, we use 5-core filtering. 
For the Yelp dataset, we follow Kheya et al.~\cite{alamgir2025unmasking} and condense the number of categories from over 300 to 21.

\begin{table}[t]
    \centering
    \caption{Details of the three datasets along with the sensitive attributes, where G=Gender, A=Age, and O=Occupation. The number after each sensitive attribute represents the number of classes for that sensitive attribute. For instance, G: 2 means gender has two classes-[male, female]. Note: in our experiments, we use binary gender, but our method can be applied to non-binary genders as well.}
    \label{tab:plain}
    \begin{tabular}{l@{\hspace{15pt}}l@{\hspace{15pt}}l@{\hspace{15pt}}l@{\hspace{15pt}}l@{\hspace{15pt}}l}
        \hline
           Name & Interactions &  Users & Items &Sensitive Attribute & Categories\\
        \hline
        ML-100K \cite{10.1145/2827872} &99,278     & 943     & 1,348 & G: 2, A: 7, O: 21 & 18 \\
        ML-1M \cite{10.1145/2827872} &   999,611  &  6,040    &  3,416& G: 2, A: 7 , O: 21 & 18\\
        Yelp \cite{mansoury2019biasdisparitycollaborativerecommendation} &  97,991   &  1,316    & 1,272 & G: 2& 21 \\
        % BookCrossing &     &      & 9424 & A \\
        \hline
    \end{tabular}
    
\end{table}

\subsection{Baselines}
\begin{figure}[t]
    \centering
    \includegraphics[width=1\textwidth]{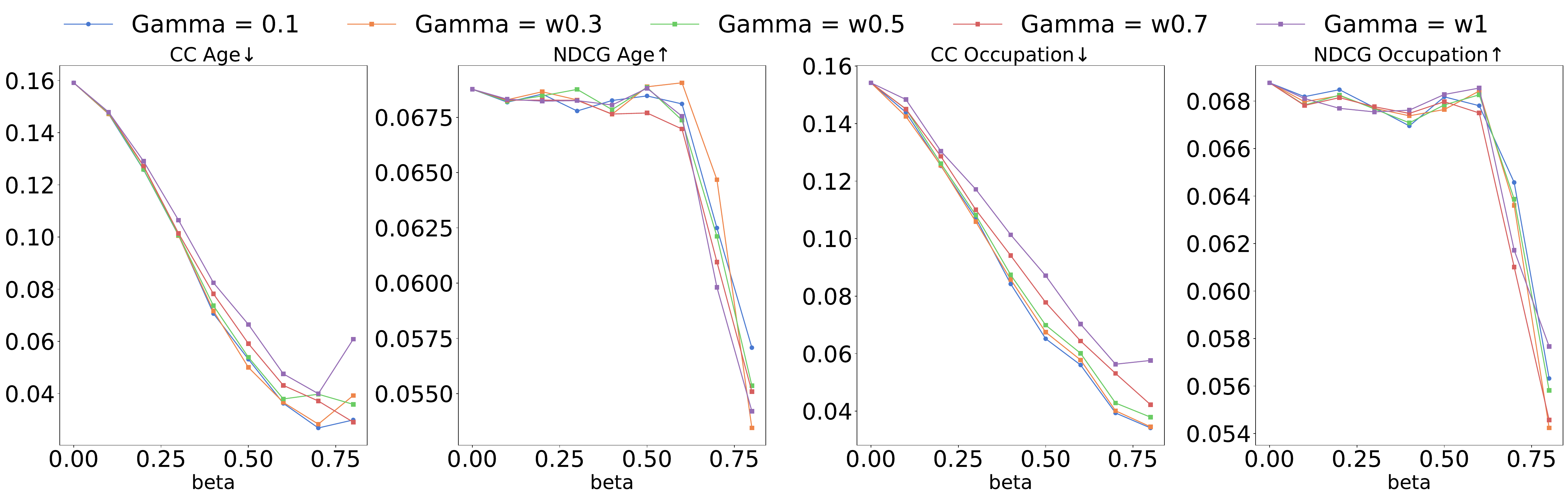}
    \caption{Impact of $\gamma$ across different $\beta$ values on NDCG and CC values for VAE-CF model for ML100K.
    % We only report these results due to space constraints. 
    }
    \label{fig:gamma}
    % \vspace{-0.5cm}
\end{figure}

As suggested by \cite{10.1145/3298689.3347058}, we evaluate our re-ranking scheme on several recommendation approaches, including traditional ones (Biased Matrix Factorization \cite{5197422} and Weighted Matrix Factorization \cite{4781121,4781145}) and deep learning-based ones (Neural Matrix Factorization \cite{10.1145/3038912.3052569} and Variational Auto Encoder Collaborative Filtering \cite{10.1145/3178876.3186150}). For all the models, we choose the best one based on the HitRatio@$20$ and NDCG@$20$ values after running them over multiple epochs for different combinations of hyperparameters. We empirically discovered that for weighing ranked items, using a gamma value of 0.1 works best in both reducing bias and minimizing performance degradation (refer to Figure \ref{fig:gamma}). For the smaller datasets, we use $N$ as the total number of items in the dataset. For the 1M datatset, $N$ is chosen to be 1000, and \(TopN_u\) for each user is the top 1,000 items for user \(u\).

\subsubsection{Bias}
For calculating bias, we extend two metrics introduced by \cite{alamgir2025unmasking}, which take into account the distribution of categories of items recommended. They were originally used to quantify gender bias in recommendations. We extend them and use them to find the sum of pair-wise differences between all user groups, as suggested by the authors, to quantify bias in multi-valued sensitive attribute groups. The first metric calculates disparity in category distributions without considering the rank, like so:
\begin{equation}
CC(c, \mathcal{U}) = \frac{1}{|\mathcal{U}|} \sum_{u \in \mathcal{U}} \frac{1}{|TopK_{u}|} \sum_{v \in TopK_{u}} \frac{C_{v,c}}{|C_{v}|}
\label{eq:bias1}
\end{equation}
The second metric scores items by discounting them based on the rank of the items in the top $k$ list. This equation is formalized as:
\begin{equation}
CDCG(c, \mathcal{U}) = \frac{1}{|\mathcal{U}|} \sum_{u \in \mathcal{U}} \frac{1}{|TopK_{u}|} \sum_{j=1}^{|TopK_{u}|} \frac{\frac{C_{v_j,c}}{|C_{v_j}|}}{\log(j + 1)}
\label{eq:bias2}
\end{equation}
The sum of pairwise differences in \textit{CC} and \textit{CDCG} values are then summed across all categories to represent the final bias values.
% \begin{equation}
    
% \end{equation}
\subsubsection{Performance}
To evaluate the performance of the models, we use HitRatio@$k$, which measures the proportion of users who get at least one relevant item recommended to them. Additionally, we use a ranking-based metric called NDCG@$k$ (Normalized Cumulative Gain) to measure the quality of the recommendations by giving higher importance to relevant items appearing higher in the list. For all our calculations, we use $k=20$.

\begin{table*}[t]
\centering
\caption{Performance and bias values across all three datasets for sensitive attributes Age (A), Gender (G), and Occupation (O).}
    \label{tab:results}
\resizebox{1\columnwidth}{!}{%
\begin{tabular}{|p{0.5cm}|llllllllllllllll|}

 \multicolumn{17}{c}{\textbf{ML100K}} 
 \\ \cline{1-17} 
% \multirow{4}{*}{\begin{tabular}[c]{@{}l@{}}\textbf{Sen.} \\ \textbf{Att}\end{tabular}} & \multicolumn{16}{c|}{}   \\ \cline{1-17} 
   & \multicolumn{8}{c|}{\textbf{MF}}                          & \multicolumn{8}{c|}{\textbf{WMF}}                             \\ \cline{2-17} 
   & \multicolumn{2}{l|}{NDCG$\uparrow$}                                          & \multicolumn{2}{l|}{HitRatio$\uparrow$}                                            & \multicolumn{2}{l|}{CC$\downarrow$}                                & \multicolumn{2}{l|}{CDCG$\downarrow$}                                & \multicolumn{2}{l|}{NDCG$\uparrow$}                                          & \multicolumn{2}{l|}{HitRatio$\uparrow$}                                            & \multicolumn{2}{l|}{CC$\downarrow$}                                & \multicolumn{2}{l|}{CDCG$\downarrow$}          \\ \cline{2-17} 
   & \multicolumn{1}{l|}{Orig.}          & \multicolumn{1}{l|}{Fair} & \multicolumn{1}{l|}{Orig.}          & \multicolumn{1}{l|}{Fair} & \multicolumn{1}{l|}{Orig.} & \multicolumn{1}{l|}{Fair} & \multicolumn{1}{l|}{Orig.} & \multicolumn{1}{l|}{Fair} & \multicolumn{1}{l|}{Orig.}          & \multicolumn{1}{l|}{Fair} & \multicolumn{1}{l|}{Orig.}          & \multicolumn{1}{l|}{Fair} & \multicolumn{1}{l|}{Orig.} & \multicolumn{1}{l|}{Fair} & \multicolumn{1}{l|}{Orignal} & Fair \\ \hline
A                                                                              & \multicolumn{1}{l|}{\multirow{3}{*}{0.0397}} & \multicolumn{1}{l|}{  0.0423}     & \multicolumn{1}{l|}{\multirow{3}{*}{0.3924}} & \multicolumn{1}{l|}{0.4210}     & \multicolumn{1}{l|}{0.1254}         & \multicolumn{1}{l|}{0.0297}     & \multicolumn{1}{l|}{0.0485}         & \multicolumn{1}{l|}{0.0107}     & \multicolumn{1}{l|}{\multirow{3}{*}{0.0372}} & \multicolumn{1}{l|}{0.0490}     & \multicolumn{1}{l|}{\multirow{3}{*}{0.3924}} & \multicolumn{1}{l|}{0.4708}     & \multicolumn{1}{l|}{0.2055}         & \multicolumn{1}{l|}{0.0302}     & \multicolumn{1}{l|}{0.0745}        &   0.0109    \\ \cline{1-1} \cline{3-3} \cline{5-9} \cline{11-11} \cline{13-17} 
G                                                                              & \multicolumn{1}{l|}{}                  & \multicolumn{1}{l|}{0.0405}     & \multicolumn{1}{l|}{}                  & \multicolumn{1}{l|}{0.4019}     & \multicolumn{1}{l|}{0.1300}         & \multicolumn{1}{l|}{0.0624}     & \multicolumn{1}{l|}{0.0507}         & \multicolumn{1}{l|}{0.0166}     & \multicolumn{1}{l|}{}                  & \multicolumn{1}{l|}{ 0.0405}     & \multicolumn{1}{l|}{}                  & \multicolumn{1}{l|}{ 0.4210}     & \multicolumn{1}{l|}{0.1888}         & \multicolumn{1}{l|}{0.0409}     & \multicolumn{1}{l|}{0.0736}        &    0.0131  \\ \cline{1-1} \cline{3-3} \cline{5-9} \cline{11-11} \cline{13-17} 
O                                                                              & \multicolumn{1}{l|}{}                  & \multicolumn{1}{l|}{0.0430}     & \multicolumn{1}{l|}{}                  & \multicolumn{1}{l|}{0.4231}     & \multicolumn{1}{l|}{0.1688}         & \multicolumn{1}{l|}{ 0.0335}     & \multicolumn{1}{l|}{0.0650}         & \multicolumn{1}{l|}{ 0.0121}     & \multicolumn{1}{l|}{}                  & \multicolumn{1}{l|}{0.0483}     & \multicolumn{1}{l|}{}                  & \multicolumn{1}{l|}{0.4719}     & \multicolumn{1}{l|}{0.2248}         & \multicolumn{1}{l|}{0.0331}     & \multicolumn{1}{l|}{0.0830}        &   0.0112   \\ \hline
\multirow{3}{*}{}                                                              & \multicolumn{8}{c|}{\textbf{VAE-CF}}                                                                                                                                                                                                                                      & \multicolumn{8}{c|}{\textbf{NeuMF}}                                                                                                                                           \\ \cline{2-17} 
                               & \multicolumn{2}{l|}{NDCG$\uparrow$}                                          & \multicolumn{2}{l|}{HitRatio$\uparrow$}                                            & \multicolumn{2}{l|}{CC$\downarrow$}                                & \multicolumn{2}{l|}{CDCG$\downarrow$}                                & \multicolumn{2}{l|}{NDCG$\uparrow$}                                          & \multicolumn{2}{l|}{HitRatio$\uparrow$}                                            & \multicolumn{2}{l|}{CC$\downarrow$}                                & \multicolumn{2}{l|}{CDCG$\downarrow$}          \\ \cline{2-17}  & \multicolumn{1}{l|}{Orig.}          & \multicolumn{1}{l|}{Fair} & \multicolumn{1}{l|}{Orig.}          & \multicolumn{1}{l|}{Fair} & \multicolumn{1}{l|}{Orig.} & \multicolumn{1}{l|}{Fair} & \multicolumn{1}{l|}{Orig.} & \multicolumn{1}{l|}{Fair} & \multicolumn{1}{l|}{Orig.}          & \multicolumn{1}{l|}{Fair} & \multicolumn{1}{l|}{Orig.}          & \multicolumn{1}{l|}{Fair} & \multicolumn{1}{l|}{Orig.} & \multicolumn{1}{l|}{Fair} & \multicolumn{1}{l|}{Orignal} & Fair \\ \hline
A                                                                              & \multicolumn{1}{l|}{\multirow{3}{*}{0.0688}} & \multicolumn{1}{l|}{0.0681}     & \multicolumn{1}{l|}{\multirow{3}{*}{0.5387}} & \multicolumn{1}{l|}{0.5451}     & \multicolumn{1}{l|}{0.1591}         & \multicolumn{1}{l|}{0.0362}     & \multicolumn{1}{l|}{0.0601}         & \multicolumn{1}{l|}{0.0135}     & \multicolumn{1}{l|}{\multirow{3}{*}{}} & \multicolumn{1}{l|}{ 0.0688}     & \multicolumn{1}{l|}{\multirow{3}{*}{}} & \multicolumn{1}{l|}{ 0.5758}     & \multicolumn{1}{l|}{0.2450}         & \multicolumn{1}{l|}{ 0.0295}     & \multicolumn{1}{l|}{0.0877}        &  0.0107   \\ \cline{1-1} \cline{3-3} \cline{5-9} \cline{11-11} \cline{13-17} 
G                                                                              & \multicolumn{1}{l|}{}                  & \multicolumn{1}{l|}{0.0682}     & \multicolumn{1}{l|}{}                  & \multicolumn{1}{l|}{0.5419}     & \multicolumn{1}{l|}{0.0516}         & \multicolumn{1}{l|}{0.0371}     & \multicolumn{1}{l|}{0.0214}         & \multicolumn{1}{l|}{0.0177}     & \multicolumn{1}{l|}{0.0708}                  & \multicolumn{1}{l|}{ 0.0688}     & \multicolumn{1}{l|}{0.5440}                  & \multicolumn{1}{l|}{ 0.5567}     & \multicolumn{1}{l|}{0.2275}         & \multicolumn{1}{l|}{ 0.0253}     & \multicolumn{1}{l|}{0.0850}        &    0.0099  \\ \cline{1-1} \cline{3-3} \cline{5-9} \cline{11-11} \cline{13-17} 
O                                                                              & \multicolumn{1}{l|}{}                  & \multicolumn{1}{l|}{0.0678}     & \multicolumn{1}{l|}{}                  & \multicolumn{1}{l|}{0.5355}     & \multicolumn{1}{l|}{0.1542}         & \multicolumn{1}{l|}{0.0561}     & \multicolumn{1}{l|}{ 0.0574}         & \multicolumn{1}{l|}{ 0.0200}     & \multicolumn{1}{l|}{}                  & \multicolumn{1}{l|}{ 0.0717}     & \multicolumn{1}{l|}{}                  & \multicolumn{1}{l|}{  0.5769}     & \multicolumn{1}{l|}{0.2402}         & \multicolumn{1}{l|}{ 0.0358}     & \multicolumn{1}{l|}{0.0878}        &    0.0136 \\ \hline

\multicolumn{17}{c}{\textbf{ML1M}}
 
% \multirow{1}{*}{\begin{tabular}[c]{@{}l@{}} \end{tabular}} & \multicolumn{16}{c|}{\textbf{ML1M}} 
% \multicolumn{17}{|c|}{\textbf{ML1M}} 
%
\\ \cline{1-17} & \multicolumn{8}{c|}{\textbf{MF}} & \multicolumn{8}{c|}{\textbf{WMF}}
 \\ \cline{2-17}  & \multicolumn{2}{l|}{NDCG$\uparrow$}                                          & \multicolumn{2}{l|}{HitRatio$\uparrow$}                                            & \multicolumn{2}{l|}{CC$\downarrow$}                                & \multicolumn{2}{l|}{CDCG$\downarrow$}                                & \multicolumn{2}{l|}{NDCG$\uparrow$}                                          & \multicolumn{2}{l|}{HitRatio$\uparrow$}                                            & \multicolumn{2}{l|}{CC$\downarrow$}                                & \multicolumn{2}{l|}{CDCG$\downarrow$}
\\ \cline{2-17}  & \multicolumn{1}{l|}{Orig.}          & \multicolumn{1}{l|}{Fair} & \multicolumn{1}{l|}{Orig.}          & \multicolumn{1}{l|}{Fair} & \multicolumn{1}{l|}{Orig.} & \multicolumn{1}{l|}{Fair} & \multicolumn{1}{l|}{Orig.} & \multicolumn{1}{l|}{Fair} & \multicolumn{1}{l|}{Orig.}          & \multicolumn{1}{l|}{Fair} & \multicolumn{1}{l|}{Orig.}          & \multicolumn{1}{l|}{Fair} & \multicolumn{1}{l|}{Orig.} & \multicolumn{1}{l|}{Fair} & \multicolumn{1}{l|}{Orignal} & Fair \\ \hline
% ########################################################################################################################
A                                                                              & \multicolumn{1}{l|}{\multirow{3}{*}{0.0363}} & \multicolumn{1}{l|}{0.0378}    & \multicolumn{1}{l|}{\multirow{3}{*}{0.3474}} & \multicolumn{1}{l|}{0.3768}      & \multicolumn{1}{l|}{0.1392}         & \multicolumn{1}{l|}{0.0537}     & \multicolumn{1}{l|}{0.0509}         & \multicolumn{1}{l|}{0.0202}     & \multicolumn{1}{l|}{\multirow{3}{*}{}} & \multicolumn{1}{l|}{0.0465}     & \multicolumn{1}{l|}{\multirow{3}{*}{}} & \multicolumn{1}{l|}{ 0.4722}     & \multicolumn{1}{l|}{0.2228}         & \multicolumn{1}{l|}{ 0.0217}     & \multicolumn{1}{l|}{0.0802}        &  0.0080   \\ \cline{1-1} \cline{3-3} \cline{5-9} \cline{11-11} \cline{13-17} 
G                                                                              & \multicolumn{1}{l|}{}                  & \multicolumn{1}{l|}{0.0377}     & \multicolumn{1}{l|}{}                  & \multicolumn{1}{l|}{0.3700}     & \multicolumn{1}{l|}{0.2016}         & \multicolumn{1}{l|}{0.0629}     & \multicolumn{1}{l|}{0.0708}         & \multicolumn{1}{l|}{0.0261}     & \multicolumn{1}{l|}{0.0428}                  & \multicolumn{1}{l|}{ 0.0455}     & \multicolumn{1}{l|}{0.4394}                  & \multicolumn{1}{l|}{ 0.4662}     & \multicolumn{1}{l|}{0.3643}         & \multicolumn{1}{l|}{ 0.0447}     & \multicolumn{1}{l|}{0.1334}        &   0.0144 \\ \cline{1-1} \cline{3-3} \cline{5-9} \cline{11-11} \cline{13-17} 
O                                                                              & \multicolumn{1}{l|}{}                  & \multicolumn{1}{l|}{0.0376}     & \multicolumn{1}{l|}{}                  & \multicolumn{1}{l|}{0.3732}     & \multicolumn{1}{l|}{0.1286}         & \multicolumn{1}{l|}{0.0597}     & \multicolumn{1}{l|}{ 0.0476}         & \multicolumn{1}{l|}{ 0.0224}     & \multicolumn{1}{l|}{}                  & \multicolumn{1}{l|}{ 0.0464}     & \multicolumn{1}{l|}{}                  & \multicolumn{1}{l|}{  0.4727}     & \multicolumn{1}{l|}{0.2374}         & \multicolumn{1}{l|}{ 0.0234}     & \multicolumn{1}{l|}{0.0867}        &    0.0103 \\ \hline
% ########################################################################################################################
\cline{2-17} & \multicolumn{8}{c|}{\textbf{VAE-CF}} & \multicolumn{8}{c|}{\textbf{NeuMF}}
 \\ \cline{2-17}  & \multicolumn{2}{l|}{NDCG$\uparrow$}                                          & \multicolumn{2}{l|}{HitRatio$\uparrow$}                                            & \multicolumn{2}{l|}{CC$\downarrow$}                                & \multicolumn{2}{l|}{CDCG$\downarrow$}                                & \multicolumn{2}{l|}{NDCG$\uparrow$}                                          & \multicolumn{2}{l|}{HitRatio$\uparrow$}                                            & \multicolumn{2}{l|}{CC$\downarrow$}                                & \multicolumn{2}{l|}{CDCG$\downarrow$}
\\ \cline{2-17}  & \multicolumn{1}{l|}{Orig.}          & \multicolumn{1}{l|}{Fair} & \multicolumn{1}{l|}{Orig.}          & \multicolumn{1}{l|}{Fair} & \multicolumn{1}{l|}{Orig.} & \multicolumn{1}{l|}{Fair} & \multicolumn{1}{l|}{Orig.} & \multicolumn{1}{l|}{Fair} & \multicolumn{1}{l|}{Orig.}          & \multicolumn{1}{l|}{Fair} & \multicolumn{1}{l|}{Orig.}          & \multicolumn{1}{l|}{Fair} & \multicolumn{1}{l|}{Orig.} & \multicolumn{1}{l|}{Fair} & \multicolumn{1}{l|}{Orignal} & Fair \\ \hline
A                                                                              & \multicolumn{1}{l|}{\multirow{3}{*}{0.0515}} & \multicolumn{1}{l|}{0.0510}     & \multicolumn{1}{l|}{\multirow{3}{*}{0.4616}} & \multicolumn{1}{l|}{0.4985}     & \multicolumn{1}{l|}{0.2076}         & \multicolumn{1}{l|}{0.0330}     & \multicolumn{1}{l|}{0.0758}         & \multicolumn{1}{l|}{0.0149}     & \multicolumn{1}{l|}{\multirow{3}{*}{}} & \multicolumn{1}{l|}{0.0470}     & \multicolumn{1}{l|}{\multirow{3}{*}{}} & \multicolumn{1}{l|}{ 0.4778}     & \multicolumn{1}{l|}{0.2554}         & \multicolumn{1}{l|}{ 0.0219}     & \multicolumn{1}{l|}{0.0928}        & 0.0081  \\ \cline{1-1} \cline{3-3} \cline{5-9} \cline{11-11} \cline{13-17} 
G                                                                              & \multicolumn{1}{l|}{}                  & \multicolumn{1}{l|}{0.0513}     & \multicolumn{1}{l|}{}                  & \multicolumn{1}{l|}{0.4884}     & \multicolumn{1}{l|}{0.2603}         & \multicolumn{1}{l|}{0.0667}     & \multicolumn{1}{l|}{0.0969}         & \multicolumn{1}{l|}{0.0309}     & \multicolumn{1}{l|}{0.0478}                  & \multicolumn{1}{l|}{ 0.0451}     & \multicolumn{1}{l|}{0.4389}                  & \multicolumn{1}{l|}{0.4684}     & \multicolumn{1}{l|}{0.4101}         & \multicolumn{1}{l|}{0.0593}     & \multicolumn{1}{l|}{0.1484}        &   0.0159 \\ \cline{1-1} \cline{3-3} \cline{5-9} \cline{11-11} \cline{13-17} 
O                                                                              & \multicolumn{1}{l|}{}                  & \multicolumn{1}{l|}{0.0519}     & \multicolumn{1}{l|}{}                  & \multicolumn{1}{l|}{0.4959}     & \multicolumn{1}{l|}{0.1514}         & \multicolumn{1}{l|}{0.0677}     & \multicolumn{1}{l|}{ 0.0542}         & \multicolumn{1}{l|}{ 0.0274}     & \multicolumn{1}{l|}{}                  & \multicolumn{1}{l|}{ 0.0468}     & \multicolumn{1}{l|}{}                  & \multicolumn{1}{l|}{ 0.4828}     & \multicolumn{1}{l|}{0.2562}         & \multicolumn{1}{l|}{ 0.0148}     & \multicolumn{1}{l|}{0.0928}        &    0.0068 \\ \hline

% ########################################################################################################################

% \multirow{1}{*}{\begin{tabular}[c]{@{}l@{}} \end{tabular}} & \multicolumn{16}{c}{\textbf{Yelp}} 
\multicolumn{17}{c}{\textbf{Yelp}}

\\ \cline{1-17} & \multicolumn{8}{c|}{\textbf{MF}} & \multicolumn{8}{c|}{\textbf{WMF}}
 \\ \cline{2-17}  & \multicolumn{2}{l|}{NDCG$\uparrow$}                                          & \multicolumn{2}{l|}{HitRatio$\uparrow$}                                            & \multicolumn{2}{l|}{CC$\downarrow$}                                & \multicolumn{2}{l|}{CDCG$\downarrow$}                                & \multicolumn{2}{l|}{NDCG$\uparrow$}                                          & \multicolumn{2}{l|}{HitRatio$\uparrow$}                                            & \multicolumn{2}{l|}{CC$\downarrow$}                                & \multicolumn{2}{l|}{CDCG$\downarrow$}
\\ \cline{2-17}  & \multicolumn{1}{l|}{Orig.}          & \multicolumn{1}{l|}{Fair} & \multicolumn{1}{l|}{Orig.}          & \multicolumn{1}{l|}{Fair} & \multicolumn{1}{l|}{Orig.} & \multicolumn{1}{l|}{Fair} & \multicolumn{1}{l|}{Orig.} & \multicolumn{1}{l|}{Fair} & \multicolumn{1}{l|}{Orig.}          & \multicolumn{1}{l|}{Fair} & \multicolumn{1}{l|}{Orig.}          & \multicolumn{1}{l|}{Fair} & \multicolumn{1}{l|}{Orig.} & \multicolumn{1}{l|}{Fair} & \multicolumn{1}{l|}{Orignal} & Fair \\ \hline

% \\ \cline{1-1} \cline{3-3} \cline{5-9} \cline{11-11} \cline{13-17} 
G                                                                              & \multicolumn{1}{l|}{0.0202}                  & \multicolumn{1}{l|}{ 0.0203}     & \multicolumn{1}{l|}{0.2660}                  & \multicolumn{1}{l|}{0.2690}     & \multicolumn{1}{l|}{0.0158}         & \multicolumn{1}{l|}{0.0202}     & \multicolumn{1}{l|}{0.0072}         & \multicolumn{1}{l|}{0.0062}     & \multicolumn{1}{l|}{0.0264}                  & \multicolumn{1}{l|}{ 0.0258}     & \multicolumn{1}{l|}{0.3590}                  & \multicolumn{1}{l|}{ 0.3389}     & \multicolumn{1}{l|}{0.0502}         & \multicolumn{1}{l|}{ 0.0247}     & \multicolumn{1}{l|}{0.0178}        &    0.0088 
\\ \hline

  \cline{2-17} & \multicolumn{8}{c|}{\textbf{VAE-CF}} & \multicolumn{8}{c|}{\textbf{NeuMF}}
 \\ \cline{2-17}  & \multicolumn{2}{l|}{NDCG$\uparrow$}                                          & \multicolumn{2}{l|}{HitRatio$\uparrow$}                                            & \multicolumn{2}{l|}{CC$\downarrow$}                                & \multicolumn{2}{l|}{CDCG$\downarrow$}                                & \multicolumn{2}{l|}{NDCG$\uparrow$}                                          & \multicolumn{2}{l|}{HitRatio$\uparrow$}                                            & \multicolumn{2}{l|}{CC}                                & \multicolumn{2}{l|}{CDCG}
 \\ \cline{2-17}  & \multicolumn{1}{l|}{Orig.}          & \multicolumn{1}{l|}{Fair} & \multicolumn{1}{l|}{Orig.}          & \multicolumn{1}{l|}{Fair} & \multicolumn{1}{l|}{Orig.} & \multicolumn{1}{l|}{Fair} & \multicolumn{1}{l|}{Orig.} & \multicolumn{1}{l|}{Fair} & \multicolumn{1}{l|}{Orig.}          & \multicolumn{1}{l|}{Fair} & \multicolumn{1}{l|}{Orig.}          & \multicolumn{1}{l|}{Fair} & \multicolumn{1}{l|}{Orig.} & \multicolumn{1}{l|}{Fair} & \multicolumn{1}{l|}{Orignal} & Fair \\ \hline
 G                                                                              & \multicolumn{1}{l|}{0.0900}                  & \multicolumn{1}{l|}{ 0.0840}     & \multicolumn{1}{l|}{0.7196}                  & \multicolumn{1}{l|}{0.6877}     & \multicolumn{1}{l|}{0.0670}         & \multicolumn{1}{l|}{0.0261}     & \multicolumn{1}{l|}{0.0235}         & \multicolumn{1}{l|}{0.0080}     & \multicolumn{1}{l|}{0.0897}                  & \multicolumn{1}{l|}{ 0.0837}     & \multicolumn{1}{l|}{0.7310}                  & \multicolumn{1}{l|}{ 0.6960}     & \multicolumn{1}{l|}{0.0617}         & \multicolumn{1}{l|}{0.0245}     & \multicolumn{1}{l|}{0.0222}        &     0.0065
\\ \hline
 
\end{tabular}}

\end{table*}

\section{Results}
\begin{figure}[t]
    \centering
    \includegraphics[width=1\textwidth]{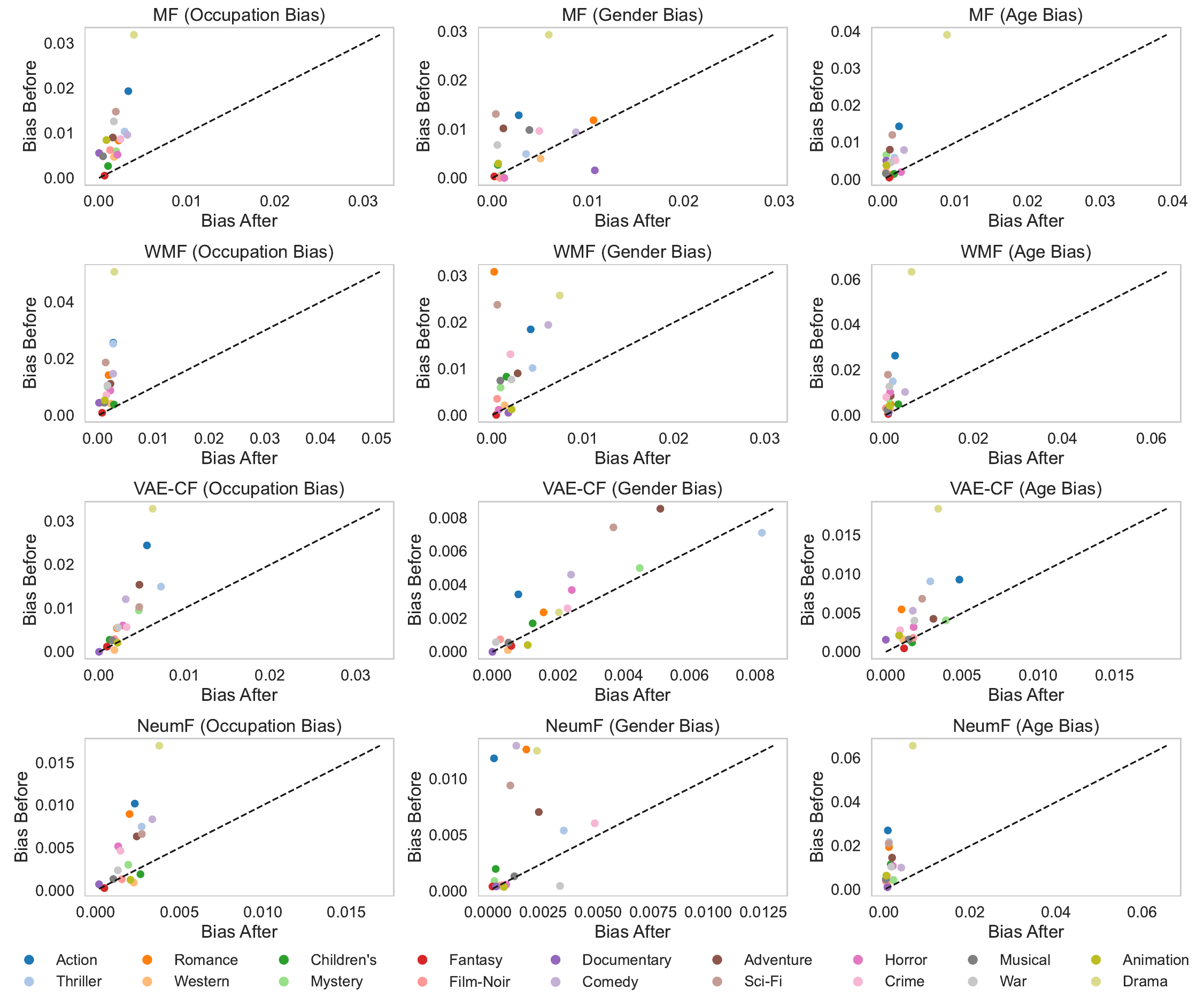}
    \caption{Comparison of bias values (CC) before and after fair re-ranking for all the models across all categories. This is only visualized for the ML100K dataset; however, other datasets have similar trends. We plot CC scores because CC and CDCG are correlated and exhibit similar patterns.
    }
    \label{fig:bias_reduction}
    % \vspace{-0.5cm}
\end{figure}

\begin{figure}[]
    \centering
    \includegraphics[width=1\textwidth]{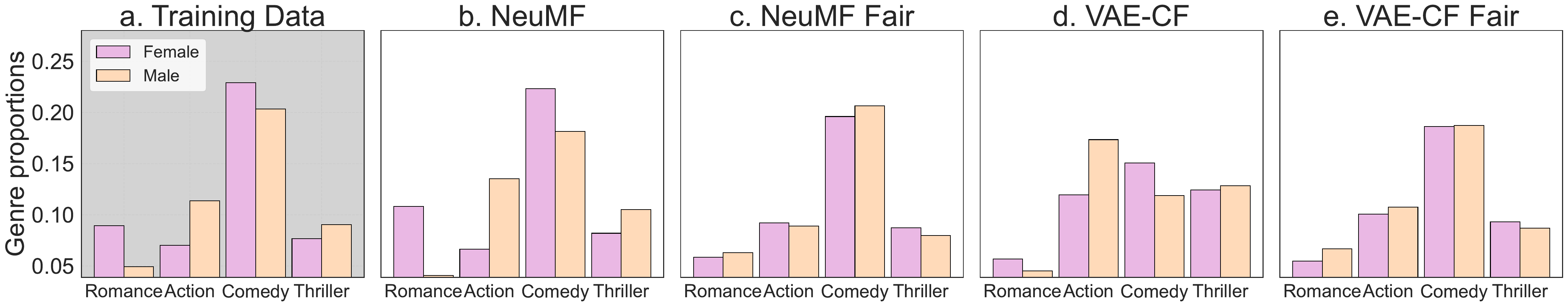}
    \caption{Comparison of recommendations before and after fair re-ranking for two of the best-performing models for the ML1M dataset across four stereotypical genres. We also show the genre proportions of the training dataset.
    }
    \label{fig:traincomparison}
    % \vspace{-0.5cm}
\end{figure}

We present the results of our experiments in Figure \ref{fig:bias_reduction}, Figure \ref{fig:abal}, Figure \ref{fig:traincomparison} and Table \ref{tab:results}. 
% We choose to visualize the bias in Figure \ref{fig:bias_reduction} by comparing the stereotypical genres: \textit{Action, Drama, Sci-Fi}, and \textit{Romance}. In our experiments, however we use all the categories. 
\subsection{Baseline Comparison}
Out of all the models, NeuMF is more prone to capturing bias relating to sensitive attributes of users. The underlying architecture of this model uses Generalized Matrix Factorization (GMF) and Multi-Layer Perceptrons
(MLP), that captures the relationships between users and the items they have interacted with. This can cause the model to capture intricate details about user preferences, which can reflect societal stereotypes \cite{alamgir2025unmasking}. VAE-CF uses probabilistic variational auto-encoders to learn user-item interactions by encoding them in latent space. While it is sensitive to capturing biases, as seen from Table \ref{tab:results}, the effect is minimal when compared to the MF-based models. Across all the models, the bias scores are higher for the larger dataset, likely due to the fact that more interactions help provide more opportunities for the model to capture the underlying biases. For performance, the deep model performs better in almost all cases, which is expected. Gender-related bias is more pronounced across all models. We believe age and occupation, may have a more subtle impact on category preferences when compared to gender. Since age and occupation have more classes than gender, the bias is more spread out across these groups making the impact more diluted. Imbalances in the interactions when considering just binary gender will stand out more, making gender bias more evident. 
\subsection{Impact of Fair-reranking}
\subsubsection{Bias}
\begin{figure}[t]
    \centering
    \includegraphics[width=1\textwidth]{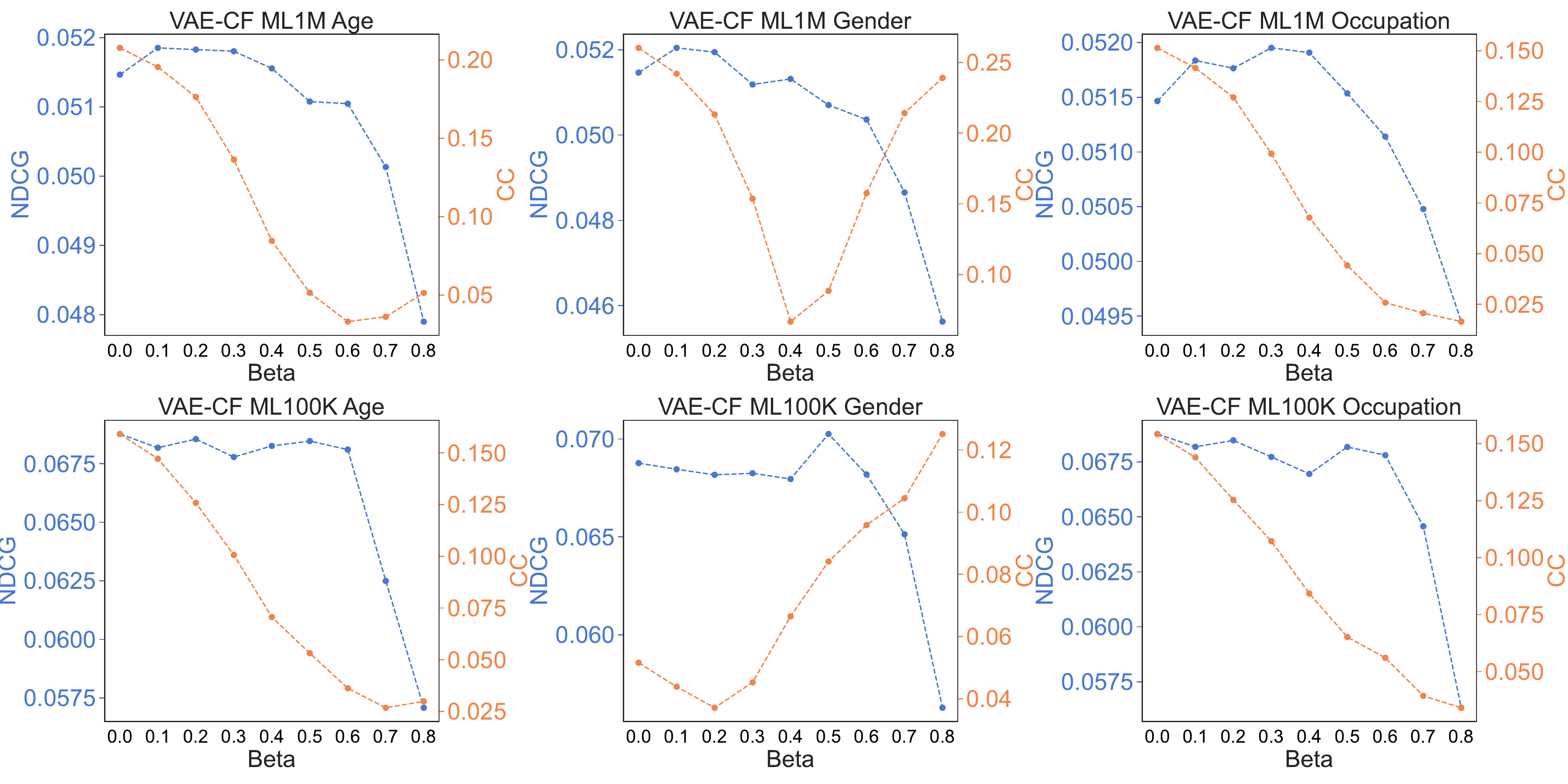}
    \caption{Impact of $\beta$ for VAE-CF model. The other models follow similar trends.
    }
    \label{fig:abal}
    % \vspace{-0.5cm}
\end{figure}

There is significant bias reduction after applying our re-ranking scheme for all models across the three sensitive attributes, as observed in Figure \ref{fig:bias_reduction}, Figure \ref{fig:traincomparison} and Table \ref{tab:results}. From Figure \ref{fig:bias_reduction}, we can observe how most points are above the y=x diagonal, verifying the reliable effectiveness of the re-ranking scheme in reducing bias. The bias mitigation works best for the NeuMF model since there's a noticeable drop in both CC and CDCG values across all the datasets for all three sensitive attributes. In Figure \ref{fig:traincomparison}, the effectiveness of our re-ranking approach is evident in plots c and e where the disparities in category distributions of recommended movies are significantly reduced compared to the baseline models (plots b and d).
We also emphasize how our re-ranking approach achieves fairness without compromising the preferences of users. For instance, both genders have strong preferences for comedies as seen in plot a, but this is not reflected in the recommendations from the VAE-CF model (plot d). However, in our fair model (plot e), the proportion of comedy movies recommended is aligned with that of the training set. The proportions for each category are calculated following the CC formula. In the case of the training set we use the historical interactions instead of the top-$k$ recommendations. While, Figure \ref{fig:traincomparison}, only shows results for ML1M dataset, readers can refer to Figure \ref{fig:compare_genre}, for ML100K dataset results. We did not include visualizations of the Yelp dataset, since they follow similar drops in bias values, like those of the other datasets. Our approach effectively minimizes discrepancies in recommended restaurants of different categories like \textit{Coffee,Tea \& Desserts} (which is more biased towards female users) and \textit{Travel \& Transportation} (which is more biased towards male users).
\\
In most cases, the performance metrics are observed to be increasing while the bias is mitigated. Theoretically, as $\beta$ increases, we would expect a decrease in bias scores and NDCG value. After a certain value of $\beta$, we would expect the bias scores to increase since the re-ranking algorithm would essentially allow the bias from CCP distribution to dominate over the actual bias. This would, in turn, start increasing the bias in the opposite direction (although we don't consider the direction of the bias because we use absolute values, we mention it here for clarity in explaining the phenomenon). To observe the influence of $\beta$ on recommendation performance and bias, we run our re-ranking algorithm for all models for $\beta$ values from 0 to 0.8, with increments of 0.1. We don't include values above 0.8 since it doesn't make sense to over-power the actual relevance scores.  As seen in Figure \ref{fig:abal}, there is a general decrease in bias values over the first few values of $\beta$. For gender, the bias starts increasing after $\beta$ is greater than a certain value (different for different models). While this trend is more prominent in the case of gender, it is also observed for age. Our intuition for a profound bias increase in gender is that the CCP distribution we are employing to fix the bias is stronger in the case of gender. So, while the distribution helps us reduce bias, as $\beta$ increases, the bias increases in the opposite direction. Again, age and occupation have more classes, and the average bias from all of these is too diluted to impact too strongly when we are using them to mitigate bias of each user's recommendations. For most models, a $\beta$ value close to 0.4-0.6, seems to work well for mitigating bias.
\subsubsection{Performance}
From Figure \ref{fig:abal}, we can observe a trend where the performance increases slightly and then decreases. While the increase in performance seems counterintuitive, this can be because the fairness term helps reduce overfitting. One way to think about this, is how the fairness term is indirectly improving the coverage of the recommendations, which in turn provides relevant items to the users. Since our approach improves the exposure of items across categories, it enhances user engagement. Additionally, the idea of improving fairness leading to improvement in performance has been observed previously by \cite{alamgir2025unmasking,10.1145/3477495.3531959,keya2020equitableallocationhealthcareresources,10.1145/3442381.3449904}. While there are some instances where there is performance drop due to increased fairness as observed in Table \ref{tab:results}, the decline is kept to a minimum.

% \usepackage{multirow}
% \usepackage{multirow}

\section{Conclusion}
In this work, we recognize the underlying issues of the current re-ranking approaches to mitigate bias in recommendations. We introduce a re-ranking scheme that reliably mitigates social bias for multi-valued user-sensitive attributes while also using item categories to ensure fine-grained treatment. Our approach is a simple yet powerful post-processing scheme to mitigate bias, which requires no modification of the model's internal parameters. We show, through extensive experiments, on three real-world datasets from multiple domains, the effectiveness of our re-ranking approach. The results show how our approach not only helps reduce bias but also preserves the quality of the recommendations, with a negligible drop in performance. We leverage the bias in the dataset to correct biased recommendations. While this works well for currently used datasets since they have historical bias, it may be less useful if future datasets evolve to be more neutral and unbiased. But we believe a dataset without bias (unless explicitly preprocessed to be fair) remains a distant possibility. While this work mainly focuses on consumers, it also implicitly accounts for the provider-side since we include item categories. In the future, we want to explicitly address the provider perspective (for instance, including item brands) to ensure a more holistic solution to social bias in recommendations. Additionally, we also intend to extend our work to address intersectional fairness for the consumers (like female and doctor).
\\
Our code, along with the processed datasets, are available here: \href{https://github.com/tahsinkheya/re_ranking_clean}{Re-ranking Code}\footnote{\url{https://github.com/tahsinkheya/re_ranking_clean}}

\section*{Acknowledgment}
This material is based upon work supported by the Air Force Office of Scientific Research under award number FA2386-23-1-4003.

%% Note that this preceding line implies that you store your BibTeX references in a file called 'mybibliography.bib'. If you instead store your references in a file with a different name, for instance 'references.bib', the preceding line should read '\bibliography{references}'. Whatever you do, DO NOT put the file name extension .bib inside the \bibliography command; this will trip up LaTeX compilers. 
%
% If you do not want to use BibTeX, you can also type up the bibliography exactly as you see fit, using the following structure:

\end{document}